\documentclass[journal]{IEEEtran}
\ifCLASSOPTIONcompsoc
  \usepackage[nocompress]{cite}
\else
  \usepackage{cite}
\fi

\usepackage{outline}
\usepackage{pmgraph}
\usepackage[normalem]{ulem}
\usepackage[utf8]{inputenc}
\usepackage{amssymb}
\usepackage{hyperref}
\usepackage{amsmath}
\usepackage{graphicx}
\usepackage{times}
\usepackage{xcolor}
\usepackage{xspace}
\usepackage[colorinlistoftodos]{todonotes} 
\usepackage{cite}
\usepackage{bm} 
\usepackage{url}

\usepackage{epstopdf}
\AppendGraphicsExtensions{.tiff}
\graphicspath{{fig/}} 

\usepackage{epsfig}
\usepackage{tikz}
\usetikzlibrary{spy}
\usepackage{algpseudocode}
\usepackage{algorithm}
\usepackage{mathrsfs}

\usepackage{nicefrac}       
\usepackage{booktabs}       

\usepackage{amsmath,graphicx,caption,mathtools,amssymb,amsthm}
\usepackage{graphicx}
\usepackage{multirow}
\usepackage{subcaption}
\usepackage{accents}

\newcommand{\add}[1] {\textcolor{blue}{#1}} 

\long\def\comment#1{} 

\newcommand{\xmath}[1] {\ensuremath{#1}\xspace}
\newcommand{\blmath}[1] {\xmath{\bm{#1}}}

\newcommand{\x}{\blmath{x}}
\newcommand{\f}{\blmath{f}}

\newcommand{\w}{\blmath{w}}

\newcommand{\norm}[1] {\xmath{\left\| #1 \right\|}}

\newcommand{\Ab}{{\blmath A}}

\newcommand{\Fb}{{\blmath F}}

\newcommand{\Ib}{{\blmath I}}

\newcommand{\Pb}{{\blmath P}}

\newcommand{\Tb}{{\blmath T}}

\newcommand{\bb}{{\blmath b}}

\newcommand{\fb}{{\blmath f}}

\newcommand{\hb}{{\blmath h}}

\renewcommand{\sb}{{\blmath s}}

\newcommand{\wb}{{\blmath w}}
\newcommand{\xb}{{\blmath x}}
\newcommand{\yb}{{\blmath y}}
\newcommand{\zb}{{\blmath z}}

\newcommand{\Nc}{\mathcal{N}}

\newcommand{\Rd}{{\mathbb R}}

\newcommand{\thetab}{{\boldsymbol {\theta}}}

\newcommand{\Ed}{{{\mathbb E}}}

\newcommand{\beq}{\begin{equation}}
\newcommand{\eeq}{\end{equation}}
\newcommand{\beqa}{\begin{eqnarray}}
\newcommand{\eeqa}{\end{eqnarray}}

\def\code#1{\texttt{#1}}
\usepackage{algorithm}
\usepackage{amsmath,graphicx,caption,mathtools,amssymb,amsthm}
\usepackage{pifont}
\newcommand{\cmark}{\ding{51}}%
\newcommand{\xmark}{\ding{55}}%
\usepackage{tikz}
\usetikzlibrary{decorations.pathreplacing,calc}
\newcommand{\tikzmark}[1]{\tikz[overlay,remember picture] \node (#1) {};}
\newcommand*{\AddNote}[4]{%
    \begin{tikzpicture}[overlay, remember picture]
        \draw [decoration={brace,amplitude=0.5em},decorate,ultra thick,black]
            ($(#3)!(#1.north)!($(#3)-(0,1)$)$) --  
            ($(#3)!(#2.south)!($(#3)-(0,1)$)$)
                node [align=center, text width=2.0cm, pos=0.5, anchor=west] {#4};
    \end{tikzpicture}
}%
\newcommand*{\AddNoteb}[4]{%
    \begin{tikzpicture}[overlay, remember picture]
        \draw [decoration={brace,amplitude=0.5em},decorate,ultra thick,black]
            ($(#3)!(#1.north)!($(#3)-(0,1)$)$) --  
            ($(#3)!(#2.south)!($(#3)-(0,1)$)$)
                node [align=center, text width=2.0cm, pos=0.5, anchor=west] {#4};
    \end{tikzpicture}
}%

\usepackage[utf8]{inputenc}
\usepackage{hyperref}
\hypersetup{
    colorlinks=true,
    linkcolor=blue,
    filecolor=magenta,      
    urlcolor=cyan,
}
\usepackage{adjustbox}

\renewcommand{\add}[1] {\textcolor{black}{#1}} 

\begin{document}
\title{MR Image Denoising and Super-Resolution \\ Using Regularized Reverse Diffusion}
\author{Hyungjin Chung,~Eun~Sun Lee$^*$,
        ~Jong~Chul~Ye$^*$,~\IEEEmembership{Fellow,~IEEE}
\thanks{H.Chung is with the Department of Bio and Brain Engineering, Korea Advanced Institute of Science and Technology (KAIST), Daejeon 34141, Republic of Korea.}
\thanks{E. S. Lee is with the Department of Radiology, Chung-Ang University Hospital, Chung-Ang University College of Medicine, 102, Heukseok-ro, Dongjak-gu, Seoul, 06973, Republic of Korea.}
\thanks{J. C. Ye is with the Kim Jaechul Graduate School of AI, the Department of Mathematical Sciences, and the Department of Bio and Brain Engineering, Korea Advanced Institute of Science and Technology (KAIST), Daejeon 34141, Republic of Korea.}
\thanks{$^*$Corresponding authors}
}

\maketitle

\begin{abstract}
Patient scans from MRI often suffer from noise, which hampers the diagnostic capability of such images. As a method to mitigate such artifact, denoising is largely studied both within the medical imaging community and beyond the community as a general subject. However, recent deep neural network-based approaches mostly rely on the minimum mean squared error (MMSE) estimates, which tend to produce a blurred output. Moreover, such models suffer when deployed in real-world sitautions: out-of-distribution data, and complex noise distributions that deviate from the usual parametric noise models.
In this work, we propose a new denoising method based on score-based reverse diffusion sampling, which overcomes all the aforementioned drawbacks. Our network, trained only with coronal knee scans, excels even on out-of-distribution {\em in vivo} liver MRI data,  contaminated with complex mixture of noise. Even more, we propose a method to enhance the resolution of the denoised image with the {\em same} network. With extensive experiments, we show that our method establishes state-of-the-art performance, while having desirable properties which prior MMSE denoisers did not have: flexibly choosing the extent of denoising, and quantifying uncertainty.
\end{abstract}


\IEEEpeerreviewmaketitle

\section{Introduction}

\IEEEPARstart{M}{agnetic} resonance imaging (MRI) is a widely used noninvasive imaging modality that can produce high resolution patient scans which aid diagnosis. Nevertheless, it is often the case where images are corrupted with complex noise, which obstructs clinicians from pointing out the details. Denoising is crucial in such cases, and for that matter, several techniques\cite{yang2015brain,manjon2008mri,awate2007feature,hanchate2020mri,veraart2016denoising,moreno2021evaluation,fadnavis2020patch2self} have been developed over the years.

Before the widespread popularity of deep learning, denoising methods such as spatial filtering\cite{tomasi1998bilateral}, transform domain filtering\cite{dabov2007image}, non local means\cite{buades2005non}, etc. were studied extensively. Nowadays, deep learning based methods are mainstream, inaugurating state-of-the-art (SOTA) over traditional methods. Denoising methods based on supervised training~\cite{zhang2017beyond,mao2016image} were the first to be developed, but these are far from being practical, since paired acquisition of clean and noisy images is rarely possible. To circumvent this difficulty, several unsupervised and self-supervised methods were proposed~\cite{lehtinen2018noise2noise,krull2019noise2void,kim2021noise2score}. Although these methods do not require paired images for training, the performance degrades when the distribution of test data deviates from the training data, or when the actual noise distribution differs from the parametric model assumption (e.g. Gaussian noise). Furthermore, most denoising methods resort to the minimum mean-squared-error (MMSE) estimate, leading to outputs that are blurrier than the noisy inputs.

These downsides are especially relevant in the context of MRI denoising. While one can simply consider thermal noise in the $k$-space acquisition as Gaussian noise,
in several cases the noise distribution is much more complex. One example of this can be seen when the data was acquired with high flip angle, reduced FOV scan. In such cases, MMSE denoisers (e.g.~\cite{lehtinen2018noise2noise}) either fail to produce feasible reconstructions due to distribution shift, or produce washed out results.

Recently, diffusion models~\cite{ho2020denoising, song2020score} have shown impressive progress in image generation~\cite{ho2020denoising, song2020score, dhariwal2021diffusion}, outperforming even the best-in-class generative adversarial networks (GAN). While diffusion models were first developed as generative models, these are now also being adopted to inverse problems including compressed sensing MRI~\cite{chung2021score, jalal2021robust, song2022solving}, CT reconstruction~\cite{song2022solving}, super-resolution~\cite{choi2021ilvr, saharia2021image, chung2022come}, and much more. Two very appealing properties of diffusion models are as follows: 1) One can acquire results from posterior sampling, rather than a single MMSE estimate. 2) They are robust to distribution shifts, and tend to generalize even to heavily out-of-distribution (OOD) test data~\cite{chung2021score}.

\begin{figure}[!hbt]
    \centering\includegraphics[width=9cm]{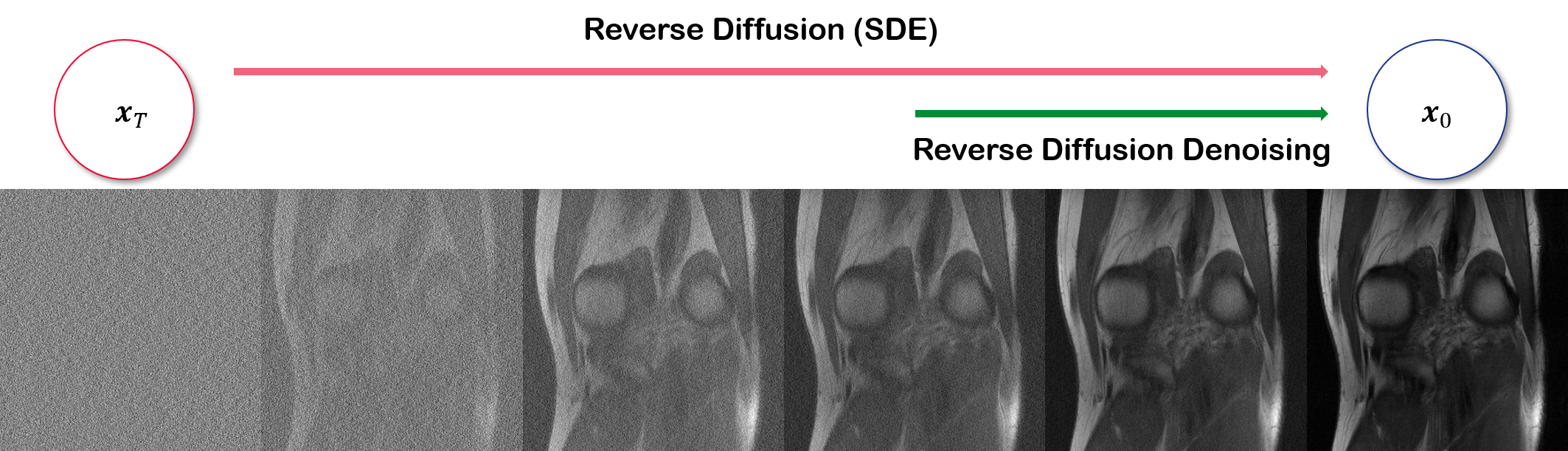}
    \caption{Overview of the proposed reverse diffusion denoising scheme. {\em Hijacking} the reverse diffusion process leads to a strong denoiser. Our model is trained on open-source knee dataset~\cite{zbontar2018fastmri}, yet is scaled to in vivo liver MRI scans.}
	\label{fig:RDD}
\end{figure}

Leveraging such intriguing properties, we propose to use a score-based diffusion model~\cite{song2020score} to solve the task of denoising. More specifically, we propose to {\em hijack} the generative process of diffusion models, not starting from pure Gaussian noise, but starting from the distribution of the noisy images. Illustrated in Fig.~\ref{fig:RDD}, this amounts to using the last few steps of the reverse diffusion, where the number of iterations can be approximated by parameter estimation methods~\cite{chen2015efficient}. Furthermore, to control the deniosing process, such that fine structures are preserved, we propose a novel low frequency constraint, which naturally connects to the recent theory of stochastic contraction in diffusion models~\cite{chung2022come}.

Moreover, we propose a method to super-resolve the denoised image with the {\em same} score function that was used to denoise the images. This immediately leads to sharper images which retain the high frequency information, which has not been reported before with any of the widely used self-supervised denoising methods.

The closest work to our proposed method is Noise2Score~\cite{kim2021noise2score}, where the authors utilize a score function to perform denoising in a single step by utilizing Tweedie's formula. Contrarily, our method {\em refines} the noisy images by  solving the reverse SDE with the score function, thereby enabling fine-grained control, and realistic results. Moreover, since our method is essentially performing posterior sampling, we can obtain multiple samples to quantify uncertainty, and to also obtain posterior mean of the distribution, as pointed out in the earlier work of CS-MRI~\cite{chung2021score}.

\section{Background}
\label{sec:background}

\subsection{Score-based diffusion model}
\label{sec:background_diffusion}

We consider the usual construction of diffusion process indexed with time $t \in [0, 1]$, which is given by $\{\x(t)\}_{t=0}^1$, with $\x(t) \in \Rd^D$. Here, $\x(0) \sim p_0$, where $p_0 = p_{\text{data}}$, and $p_1$ approximates isotropic Gaussian distribution. In other words, the data distribution is transformed to a tractable Gaussian as $t \mapsto 1$. To formally represent this, we first introduce the following forward SDE
\begin{equation}\label{eq:fsde}
    d\x = \fb(\x, t)dt + g(t)d\w,
\end{equation}
where $\f$ is a linear drift coefficient, $g$ is a scalar diffusion coefficient, and $\w$ is a standard $D$-dimensional Wiener process. In our case, we define $\f, g$ to be
\begin{equation}\label{eq:vesde}
\fb = 0,\quad g = \sqrt{\frac{d[\sigma^2(t)]}{dt}},
\end{equation}
with 
\begin{equation}\label{eq:sigmat}
    \sigma(t)= \sigma_{\text{min}}\left(\frac{\sigma_{\text{max}}}{\sigma_{\text{min}}}\right)^t.
\end{equation}
This choice is called variance exploding SDE (VE-SDE)~\cite{song2020score}, and is widely used due to its impressive sample quality. By Anderson's theorem~\cite{anderson1982reverse}, we subsequently have the following reverse SDE
\begin{align}\label{eq:rsde}
    d\xb &= [\fb(\xb, t) - g(t)^2 \underbrace{\nabla_\xb \log p_t(\xb)}_{\text{score function}}]dt + g(t) d\bar{\wb} \\
    &= \frac{d[\sigma^2(t)]}{dt} \underbrace{\nabla_\xb \log p_t(\xb)}_{\text{score function}}dt + \sqrt{\frac{d[\sigma^2(t)]}{dt}} d\bar{\wb} \notag,
\end{align}
where the differential $dt$ indicate time running backwards, and $\bar{\wb}$ is again a standard $D$-dimensional Wiener process running backwards in time.

Here, we see that the reverse SDE involves the score function (i.e. gradient of the log probability). In order to estimate the score function, denoising score matching~\cite{vincent2011connection} objective is used, which sidesteps the difficulties, and makes the training easily scalable to higher dimensions~\cite{song2019generative}. More specifically, we train a time-conditional neural network $\sb_\theta(\x, t)$ with the following objective:
\begin{align}\label{eq:score_cost}
    \min_{\thetab} \Ed_{t \sim U(0, 1)} \Big[ \lambda(t) &\Ed_{\xb(0)} \Ed_{\xb(t)|\xb(0)}\Big[ \\ \notag
    &\norm{\sb_{\thetab} (\xb(t), t) - \nabla_{\xb} \log p_{0t} (\xb(t) | \xb(0))}_2^2 \Big]
    \Big],
\end{align}
where $U(0,1)$ denotes the uniform distribution in the interval $[0,1]$.
Here, we use the fact that the perturbation kernel $\nabla_{\xb} \log p_{0t} (\xb(t) | \xb(0))$ is always Gaussian when $\f$ is chosen to be affine, which is trivially true in our case. Hence, the derivative with respect to $\x$ can be computed analytically, which makes the training procedure simple and scalable.

Once the network is optimized via \eqref{eq:score_cost}, one can plug in the trained score function to numerically solve \eqref{eq:rsde}. This can be done by e.g. Euler-Maruyama discretization~\cite{song2020score}, or higher order methods~\cite{jolicoeur2021gotta}. On the other hand, one can use predictor-corrector (PC) samplers~\cite{song2020score}, which alternate between numerical SDE solver and Langevin dynamics~\cite{parisi1981correlation}.

\subsection{Come-Closer-Diffuse-Faster (CCDF) \cite{chung2022come}}
A downside of diffusion models is that they are very slow. This is because one needs to start sampling from totally random Gaussian noise, and thus few thousand  passes through the neural network is necessary to achieve optimal results. Recently, Chung {\em et al.}~\cite{chung2022come} showed that such is not necessarily the case when solving inverse problems with diffusion models. Rather than starting from Gaussian noise, one can forward diffuse the initial corrupted image (measurement), and only use the last few steps of reverse diffusion iteration to arrive at the final reconstruction, given that one uses a proper non-expansive mapping as a data consistency imposing step~\cite{bauschke2011convex}. 

This intriguing behavior is originated from the key observation that the score-based reverse diffusion process itself  is a
stochastic contraction mapping so that as long as the data consistency imposing mapping is non-expansive, the alternating applications of the
reverse diffusion and data consistency term results in a stochastic contraction to a fixed point \cite{chung2022come}.
This theory is ideally suitable for the  inverse problems, as the steps which impose data fidelity can be easily cast as non-expansive mapping \cite{chung2022come}. 
In Section~\ref{sec:main_contributions}, we discuss ways to regularize the reverse diffusion process for our purpose by applying such non-expansive mapping.

\subsection{Noise2Score}
\label{sec:background_noise2score}
Having access to estimated score also endows us with other capabilities. Suppose that we have the noising process as
\begin{equation}\label{eq:noise_model}
    \yb = \xb + \wb,\, \wb \sim \Nc(0, \eta^2 \Ib),
\end{equation}
where $\yb \in \Rd^D$ is the noisy measurement, $\xb \in \Rd^D$ is the latent clean image, and $\wb \in \Rd^D$ represents the noise vector sampled from $\Nc$. Here, $\Nc(0, \eta^2 \Ib)$ denotes zero-mean Gaussian with variance $\eta^2$. From this model, in order to estimate the clean image $\xb$ from the noisy image $\yb$, Tweedie's formula states that the posterior mean of $\xb$ given $\yb$ is \cite{kim2021score}
\begin{equation}\label{eq:tweedie}
    \Ed[\xb|\yb] = \yb + \eta^2 \nabla_{\yb} \log p(\yb).
\end{equation}
In another word, we can find the minimum mean square error (MMSE) estimator, given the correct score function of $\yb$. It was shown by the authors of~\cite{kim2021score} that one can apply similar procedures to the general noise distributions that follow the exponential family (e.g. Poisson, Gamma).

\section{Main Contributions}
\label{sec:main_contributions}

Throughout the paper, 
we discretize the time index $t$ into $N$ intervals uniformly. In this case, our noise schedule of $\sigma(t)$ in \eqref{eq:sigmat}  and the discretized sample
become
\begin{equation}
    \sigma_i := \sigma_{\text{min}}\left(\frac{\sigma_{\text{max}}}{\sigma_{\text{min}}}\right)^{\frac{i-1}{N-1}},\quad \xb_i := \left. \xb(t)\right|_{t=\frac{i-1}{N-1}}.
\end{equation}

\subsection{Score function as denoiser}

Consider a noisy image $\xb(t)$, defined by the forward diffusion process, which essentially samples from $p(\xb(t)|\xb(0))$:
\begin{equation}
\label{eq:conditional_0t}
    \xb(t) = \xb(0) + \wb, \, \wb \sim \Nc(0, \sigma(t)^2\Ib),
\end{equation}
where $\sigma(t)$ is defined in Eq.~\eqref{eq:sigmat}. We see that this is identical to \eqref{eq:noise_model} when $\eta = \sigma(t)$. Moreover, suppose that $\sb_\theta$ is expressive enough and trained to optimality such that $\sb_{\theta^*} = \nabla_{\xb_t}p_{t}$, i.e.
\begin{align*}
    \sb_{\theta^*} (\xb(t), t) &= \nabla_{\xb_t}\left(-\frac{\|\xb(t) - \xb(0)\|^2}{2\sigma(t)^2}\right) 
    = - \frac{\xb(t) - \xb(0)}{\sigma(t)^2}.
\end{align*}
Subsequently, using Tweedie's formula in \eqref{eq:tweedie} gives us
\begin{align}
    \Ed[\xb(0)|\xb(t)] &= \xb(t) + \sigma(t)^2 \sb_{\theta^*}(\xb(t), t) \notag \\
    &= \xb(t) + \sigma(t)^2 \left(-\frac{\xb(t) - \xb(0)}{\sigma(t)^2}\right) = \xb(0). \label{eq:tweedie2}
\end{align}
With \eqref{eq:tweedie2}, we again confirm that it is  possible to use the estimated score function trained under the score-SDE framework~\cite{song2020score} to estimate the denoised image of $\xb(t)$. The only requirements are that the noise variance $\sigma(t)$ lies within the bound of the noise schedule for training, and that we can correctly estimate $\sigma(t)$.
In this case, Eq.~\eqref{eq:tweedie2} provides us with a method to denoise $\xb(t)$ in a {\em single} step. 

However, the single-step MMSE denoising often produce blurred results, which is undesirable since this may raise the difficulty of clinical decision making. The situation gets severer as the degree of noise scale increase. Indeed, this is a problem not only for denoising with score function, but also for most of the recent self-supervised  denoisers~\cite{kim2021score,huang2021neighbor2neighbor,lehtinen2018noise2noise,krull2019noise2void} relying on a single-step estimate.

This suggests that fine-grained control and posterior sampling of the denoising estimate would be beneficial for obtaining sharp and high fidelity denoised images. Fortunately, given the generative reverse SDE in \eqref{eq:rsde}, we can deal with the difficulty, which we explore in the following section.

\subsection{Reverse diffusion as denoiser}

Recall that the reverse SDE in \eqref{eq:rsde} can be numerically solved (integrated) by methods such as Euler-Maruyama discretization. In the case of solving reverse SDE for VE-SDE, this amounts to iteratively applying
\begin{align}
\label{eq:solver}
    &\text{Solver}(\xb_{i+1},\zb) := \notag \\
    &\xb_{i+1} + (\sigma_{i+1}^2 - \sigma_i^2)\sb_\theta(\xb_{i+1},\sigma_{i+1}) + \sqrt{\sigma_{i+1}^2 - \sigma_i^2}\zb,
\end{align}
as shown in Algorithm~\ref{alg:rsde}. Here, we have written down a single Euler-Maruyama step~\cite{sarkka2019applied} for simplicity, but application of other solvers is straightforward.
The solver in Algorithm~\ref{alg:rsde} can be thought of as generating a {\em sample path}~\cite{sarkka2019applied}, given some initial random noise sample $\xb_N \sim p_t(\xb)$. By doing so, one would be able to achieve a sample from $p_0(\xb)$, given that the score function was properly trained.

\begin{algorithm}[!hbt]
\caption{Numerical solving of \eqref{eq:rsde}}
\begin{algorithmic}[1]
\Require $\{\sigma_i\}_i$
\State $\xb_N \sim \Nc({\bf 0}, \sigma_{\text{max}}^{2} \Ib)$
\For{$i = N-1:0$} \do \\
\State $\zb \sim \Nc({\bf 0}, \Ib)$
\State $\xb_i \gets \text{Solver}(\xb_{i+1},\zb)$
\EndFor
\State \textbf{return} $\xb_0$
\end{algorithmic}\label{alg:rsde}
\end{algorithm}

On the contrary, let us define a noisy image $\xb_{N'}:=\xb({t'})$, where $N' := Nt'$, and $t' \in [0, 1]$. In this case, we can consider that the given noisy image is a sample from $p_{t'}(\xb)$. When we know the noise level $\sigma_{\text{true}}$ of $\xb({t'})$, we can specify the value of $t'$ by $t' = \sigma^{-1}(\sigma_{\text{true}})$. Then, we can generate a sample path starting from the initial value $\xb({t'})$, which would amount to iterating the solver for $N'$ steps, rather than $N$.

Although in the easier case when we know a priori the amount of noise, we can simply choose $t'$ such that $\sigma(t')$ matches the known noise variance,
 in most practical situations, this is not possible and one should resort to some surrogate method to estimate the noise parameters. For that matter, we propose to use a non-parameteric estimation method based on eigenvalue analysis of the covariance matrix~\cite{chen2015efficient}, which we denote in Algorithm~\ref{alg:rdd} with $E$. Note that this is simply a design choice, and one can use other estimators as well~\cite{pyatykh2012image,liu2012additive,liu2013single}. One should note that the noise distribution may deviate from the Gaussian distribution. However, this is not a big problem thanks to the generalization capacity of diffusion models~\cite{chung2021score}.

\begin{algorithm}[!hbt]
\caption{Reverse diffusion denoiser}
\begin{algorithmic}[1]
\Require $\{\sigma_i\}_i$, $\xb_{N'}, E, \alpha$
\State $\sigma_{est} = E(\xb_{N'})$ \Comment{Noise level estimation}
\State $t' = \sigma^{-1}(\sigma_{est})$
\State $N' = \alpha t'N$
\For{$i = N'-1:0$} \do \\
\State $\zb \sim \Nc({\bf 0}, \Ib)$
\State $\xb_i \gets \text{Solver}(\xb_{i+1},\zb)$
\EndFor
\State \textbf{return} $\xb_0$
\end{algorithmic}\label{alg:rdd}
\end{algorithm}

Another hyper-parameter to endow further flexibility is $\alpha$ in Algorithm~\ref{alg:rdd}. According to the circumstances, a clinician might either have to apply denoising aggressively such that one can maximally reduce the noise level, or apply it mildly so that one can best preserve the original details. Hence, one can decide on the value $\alpha$, or simply apply Algorithm~\ref{alg:rdd} with different values of $\alpha$ multiple times, to choose the best-performing value. We discuss this interesting aspect of the proposed method further in Section~\ref{sec:flexibility}.

\subsection{Low frequency regularizer}

Ideally, one may use Algorithm~\ref{alg:rdd} and achieve denoised results. However, due to the inherent stochasticity, simply applying Algorithm~\ref{alg:rdd} may result in altered structure. Especially in medical imaging, this could lead to catastrophic outcome. In order to prevent such outcome, we propose a low frequency regularization scheme.

Specifically, we focus on the fact that noise components are mostly focused on the high frequency part of the $k$-space of the given image. Hence, it would be reasonable to keep the low frequency component intact, and focus mainly on correcting the high frequency component. Formally, the regularization step can be written as
\begin{align}
\label{eq:lf_reg}
    \xb_{i+1} = \lambda\Fb^{-1}\Pb_{\Omega}\Fb \xb_{N'} + (1 - \lambda)\xb_i,
\end{align}
where $\Fb \in \Rd^{D\times D}$ denotes the discrete Fourier transform matrix, $\Pb_{\Omega} \in \Rd^{D\times D}$ is a diagonal sub-sampling matrix with ones at the low frequency region $\Omega$, and $\lambda \in [0, 1]$ is a hyper-parameter which puts emphasis on the regularization.
Note that \eqref{eq:lf_reg} can be written as
\begin{align*}
    \xb_{i-1} = \Tb(\xb_i):= \Ab\xb_i + \bb,
\end{align*}
where
$\Ab =  (1 - \lambda)\Ib$ and $\bb =  \lambda\Fb^{-1}\Pb_{\Omega}\Fb \xb_{N'}$.
Hence, $\Tb$ is nonexpansive  since we have    $\|\Ab\|=\|(1 - \lambda)\Ib\| \leq 1$.
Therefore, we can see that \eqref{eq:lf_reg} not only regularizes the structure to be unaltered, but it also accelerates the contraction to a feasible solution, as proved in~\cite{chung2022come}.

\begin{figure}[!hbtb]
    \centering\includegraphics[width=6.5cm]{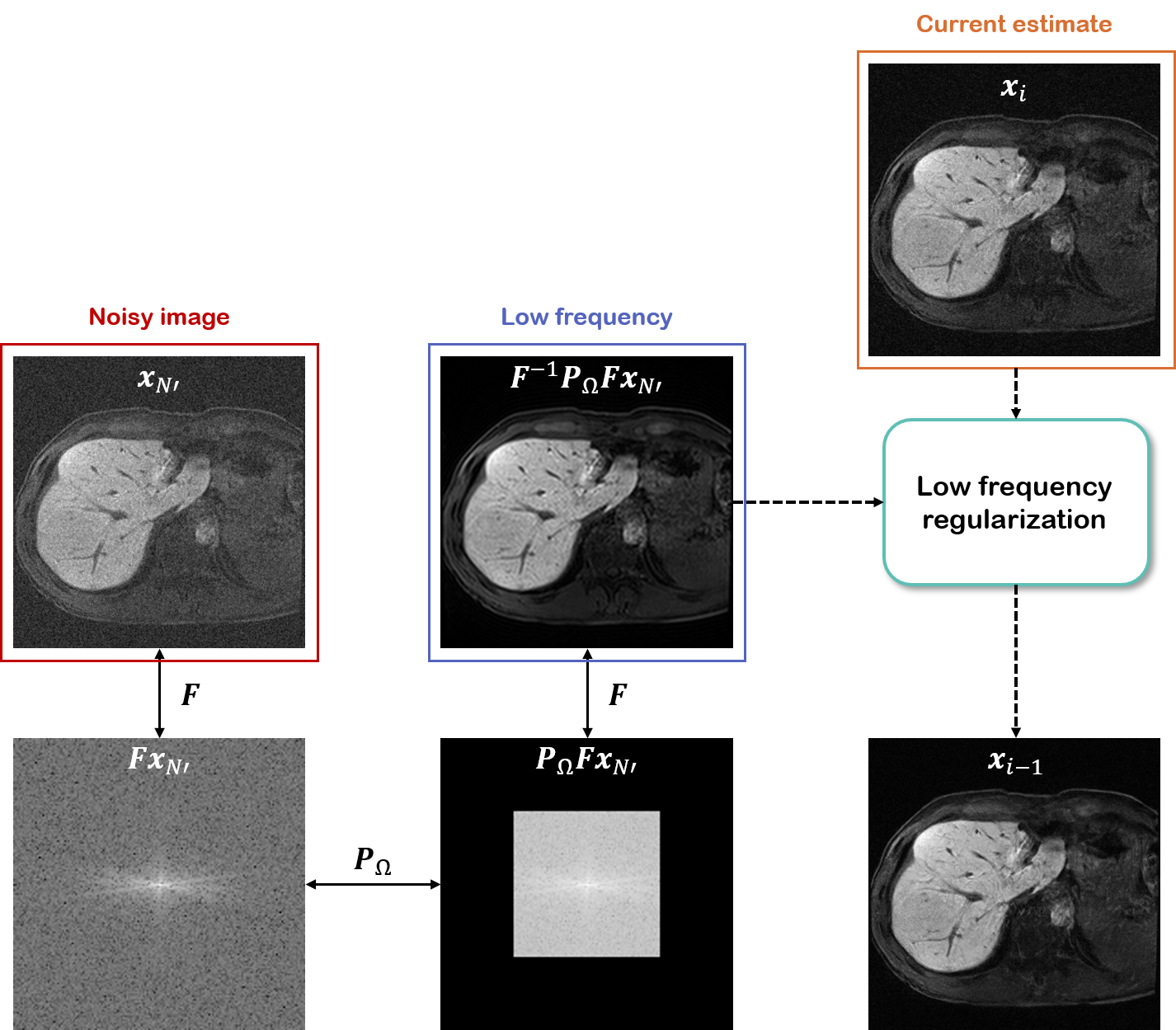}
    \caption{Illustration of low frequency regularization.}
    \vspace*{-0.5cm}
	\label{fig:lf_reg}
\end{figure}

 An illustration of the regularization is depicted in Fig.~\ref{fig:lf_reg}. Essentially, the low frequency part from the initial noisy image $\xb_{N'}$ is softly injected at every iteration in order to keep the important structure intact.

\subsection{Post-hoc super-resolution}
It is often the case that an image that we wish to see not only is corrupted with noise, but also has low resolution. In such cases, it would be ideal if we could apply super-resolution to the denoised image, since now the noise component would not be amplified. Unfortunately, conventional denoisers are devoid of such ability. In contrast, it has been studied that a single score function can be used to solve multiple inverse problems, when the right consistency term is applied~\cite{chung2022come}. 

Here, we show that we can design a super-resolution algorithm that is applied once the denoising is complete. 
Specifically, we define a blur kernel $\hb_D$ which is defined by  successive applications of the downsampling filter by
a factor $D$, and upsampling filter by a factor $D$.
This can be represented as a matrix multiplication:
$\Pb \x' :=\hb_D \ast \x'$,
where $\x'$ denotes intermediate estimate from the reverse diffusion.
Then, we use the following data consistency iteration:
\begin{align}\label{eq:super}
\hat\x_i=  \Tb(\hat\x_i'):= {(\Ib-\Pb)\hat\x_i'}  + {{{\x}_0}},
\end{align}
where $\hat\x_i'$ is the current estimate from the diffusion, and
$\x_0$ is the denoised image from the denoising step.
As shown in \cite{chung2022come},  we can see that  $\Tb$  in  \eqref{eq:super} is non-expansive
 for the normalized filter $\hb_D$.
Therefore, one can alternately apply the reverse diffusion and data consistency in a recursive manner to achieve a fixed point through stochastic contraction:
\begin{align*}
    \hat\xb'_i &= \hat\xb_{i+1} + (\sigma_{i+1}^2 - \sigma_{i}^2) \sb_{\theta}(\hat\xb_{i+1}, \sigma_{i+1}) + \sqrt{\sigma_{i+1}^2-\sigma_i^2}\zb\\
    \hat\xb_i &= (\Ib - \Pb)\hat\xb'_i + \xb_0. 
\end{align*}

Summing everything up, we arrive at Algorithm~\ref{alg:r2d2_sr}, which we call R2D2+\footnote{Short for \textbf{R}egularized \textbf{R}everse \textbf{D}iffusion \textbf{D}enoiser \textbf{+} SR}. Note that we have initialized the SR process with a forward diffusion. i.e.
\begin{equation}
    \hat\xb_M = \xb_0 + \sigma_M\zb.
\end{equation}
Hence, rather than starting from random Gaussian noise as in~\cite{choi2021ilvr}, one can start from $\hat\xb_M$, and use small number of iterations to achieve reconstruction, as introduced as CCDF strategy in~\cite{chung2022come}. Accordingly, both the denoising and the SR steps of R2D2+ requires few tens of iterations, as opposed to other diffusion models which require few thousand steps of iterations~\cite{song2020score,choi2021ilvr,ho2020denoising}.

%
%

\begin{algorithm}[!hbt]
\caption{Regularized reverse diffusion denoiser + SR}
\begin{algorithmic}[1]
\Require $\{\sigma_i\}_i$, $\xb_{N'}, E, M, \alpha$
\State $\sigma_{est} = E(\xb_{N'})$ \Comment{Noise level estimation}
\State $t' = \sigma^{-1}(\sigma_{est})$
\State $N' = \alpha t'N$
\For{$i = N'-1:0$} \do  \\
\State $\zb \sim \Nc({\bf 0}, \Ib)$ \tikzmark{top}  
\State $\xb_i \gets \text{Solver}(\xb_{i+1}, \add{\zb})$ 
\State $\xb_{i} \gets \lambda\Fb^{-1}\Pb_{\Omega}^{-1}\Pb_\Omega\Fb\xb_{N'} + (1 - \lambda)\xb_i$ \tikzmark{bottom}\tikzmark{right}
\EndFor
\State $\zb \sim \Nc({\bf 0}, \Ib)$
\State $\hat\xb_M \gets \xb_0 + \sigma_{M}\zb$
\For{$j = M-1:0$} \do  \\
\State $\zb \sim \Nc({\bf 0}, \Ib)$ \tikzmark{top2}  
\State $\hat\xb_j \gets \text{Solver}(\hat\xb_{j+1}, \add{\zb})$
\State $\hat\xb_j = (\Ib - \bm\Phi)\hat\xb'_j + \xb_0$ \tikzmark{bottom2}
\EndFor
\State \textbf{return} $\hat\xb_0$
\AddNote{top}{bottom}{right}{Denoiser}
\AddNoteb{top2}{bottom2}{right}{SR}
\end{algorithmic}\label{alg:r2d2_sr}
\end{algorithm}


\section{Methods}
\label{sec:methods}

\subsection{Experimental Data}

For our denoising experiment, we do not have access to ground-truth clean liver MR scans. Nevertheless, thanks to the superb generalization capacity of score-based diffusion models~\cite{chung2021score,jalal2021robust}, we can train our network with open-sourced data such as the fastMRI~\cite{zbontar2018fastmri} knee datset\footnote{\href{https://fastmri.org/}{https://fastmri.org/}}. Specifically, we train the network as advised in~\cite{chung2021score}, using 320$\times$320 sized fully-sampled single coil MRI magnitude images. Since the data is a simulated single-coil measurement, the ground truth images itself are contaminated with various sources of noise. However, this also is not a problem, since it was shown that one can also train a score function with noisy images~\cite{kim2021noise2score}.

{The data for testing was collected from Chungang University Medical center, with the Siemens Skyra scanner. The testing data consists of 28 3-D volumes, acquired via high flip angle CAIPIRINHA~\cite{yu2013clinical}\footnote{\textbf{C}ontrolled \textbf{A}liasing \textbf{I}n \textbf{P}arallel \textbf{I}maging \textbf{R}esults \textbf{I}n \textbf{H}igher \textbf{A}cceleration}, with reduced FOV and thin slice thickness (1.5 mm) to capture small lesions. Specific parameters for the dataset are listed in TABLE~\ref{tab:data_spec}.}

\begin{table}[!bt]
\begin{center}
	\resizebox{0.4\textwidth}{!}{
    \begin{tabular}{l|c}
        \hline
    Parameter     & value \\
    \hline\hline
    Scanner     & Siemens 3T Skyra scanner\\
    Contrast & T1W fast suppressed (FS)\\
    Sequence & Gradient Echo (GRE)\\
    Time of Repetition(TR) [ms] & 6.42\\
    Time of Echo (TE) [ms] & 2.46 \\
    Echo Train Length (ETL) & 2 \\
    Matrix size & 256 $\times$ 256 \\
    Resolution [mm$^3$] & 1.0 $\times$ 1.0 $\times$ 1.5\\
    \hline
    \end{tabular}}
\caption{Specifications for the liver MRI test dataset.}
\label{tab:data_spec}
\end{center}
\vspace*{-0.5cm}
\end{table}

\subsection{Details of implementation}

We train our score function $\sb_\theta$ based on VE-SDE cost~\cite{song2020score}, formally written as
\begin{align}\label{eq:score_cost_VESDE}
    \min_{\thetab} &\Ed_{t \sim U(\epsilon, 1)} \Ed_{\xb(0) \sim p_0} \Ed_{\xb(t) \sim \Nc(\xb(0),\sigma^2(t)\Ib)}\Big[ \\ \notag
    &\lambda(t)\norm{\sb_{\thetab} (\xb(t), t) - \frac{\xb(t) - \xb(0)}{\sigma^2(t)}}_2^2
    \Big],
\end{align}
with $\lambda = \sigma^2(t)$, such that one achieves the maximum likelihood training objective of~\cite{song2021maximum}. To avoid numerical issues from the unbounded values of score~\cite{kim2021score}, we set $\epsilon = 10^{-5}$.

In the function $\sigma(t)$, used in the diffusion function of the VE-SDE, we need to specify $\sigma_{\text{min}}$ and $\sigma_{\text{max}}$. Following the parameters of~\cite{chung2021score}, we set $\sigma_{\text{min}} = 0.01$, $\sigma_{\text{max}}$ = 378. We use a batch size of 2, and train the network using Adam~\cite{kingma2015adam} optimizer with linear warm-up schedule. Specifically, we linearly increase the learning rate for the first 5000 steps of optimization, reaching \code{2e-4} at the end of warmup. Learning rate remains static after the warmup. Exponential moving average of 0.9999 is applied to the parameters, and the training took about 3 weeks with 2$\times$RTX 2080Ti GPUs. We use \code{ncsnpp}~\cite{song2020score} as the neural network architecture for modeling $s_\theta$, which conditions the network on $t$ with Fourier features~\cite{tancik2020fourier}. For the numerical SDE solver, we use predictor-corrector (PC) sampler proposed in~\cite{song2020score}. The parameter for low frequency regularization is set to $\lambda = 0.005$. We set a constant $\alpha = 0.2$ unless specified otherwise. We use the official implementation of~\cite{chen2015efficient}, with the default parameter (patch size of 8) for the estimation of noise variance.

\subsection{Comparison methods}

We compare the proposed method with the state-of-the-art (SOTA) methods. First, to compare against a representative non-deep learning method, we use Block Matching \& 3D filtering (BM3D). BM3D uses collaborative filtering with similar blocks extracted from the same image. In order to denoise an image, BM3D requires the level of noise variance, which we estimate using the same method that was used for the noise estimation of the proposed method~\cite{chen2015efficient}. We use the official python implementation\footnote{\href{https://pypi.org/project/bm3d/}{https://pypi.org/project/bm3d/}} with the default settings, which utilize both Wiener filtering and hard thresholding.

For deep learning methods, we cannot use the supervised learning approaches as there are no ground-truth data available. Moreover,
{cycleGAN-based approaches~\cite{gu2021cyclegan,chung2021simultaneous,kwon2021cycle} are not feasible in this case, as we do not have access to clean liver data set albeit unmatched. }
Therefore, we compare with self-supervised learning approaches, such as Noise2Noise (N2N)~\cite{lehtinen2018noise2noise}, Neighbor2Neighbor (Nei2Nei)~\cite{huang2021neighbor2neighbor}, and Noise2Score (N2Score)~\cite{kim2021noise2score}. In order to train both N2N and Nei2Nei methods, we use random gaussian noise of scale uniformly ranging from $\sigma = 5.0$ to $\sigma = 50.0$, with the default settings as advised in the original works. For N2Score, we only need a pre-trained score function, and we use the same score function that was used for  the proposed method.

\subsection{Quantitative evaluation and statistical analysis}

The primary objective of the proposed method is to increase the signal-to-noise ratio while also boosting the resolution of the images. In order to quantify the improvement, we use two standard metrics: signal-to-noise ratio (SNR), and contrast-to-noise ratio (CNR). Specifically, we calculate the metrics with the following equation:
\begin{equation}
    SNR = \frac{\mu(\xb_{[s]})}{\varsigma(\xb_{[s]})}, \quad
    CNR = \frac{\mu(|\xb_{[s]} - \xb_{[b]}|)}{\varsigma(\xb_{[b]})}
\end{equation}
where $\xb_{[s]}, \xb_{[b]}$ refers to the circular masked region of interest (ROI) where the signal is dominant, and the circular masked ROI where the background resides, respectively. Moreover, $\mu$ refers to the mean value, and $\varsigma$ indicates the signal standard deviation.

When calculating the SNR, we manually selected 8 distinct regions per volume, corresponding to the 8 different liver segments~\cite{sibulesky2013normal}. Circular disk with the radius of 10 mm was chosen to mask out the ROIs. For the calculation of CNR, we focused on 2 different regions per volume, in which we select two major vessels in the left liver, and the right liver, respectively.
As stated in the previous sections, we do not have ground truth data to acquire standard metrics such as peak signal-to-noise ratio (PSNR) or structural similarity index (SSIM).

In order to confirm the statistical relevance of the reported metrics, we conducted RM-ANOVA\footnote{\textbf{R}epeated \textbf{M}easures \textbf{AN}alysis \textbf{O}f \textbf{V}ariance}, considering results obtained with each method as groups. All calculations were performed with the MedCalc\footnote{\href{https://www.medcalc.org/}{https://www.medcalc.org/}} software. Bonferroni correction was performed post-hoc to RM-ANOVA to correct for errors that arise when dealing with more than 2 groups.

\begin{figure*}[!hbt]
    \centering\includegraphics[width=18cm]{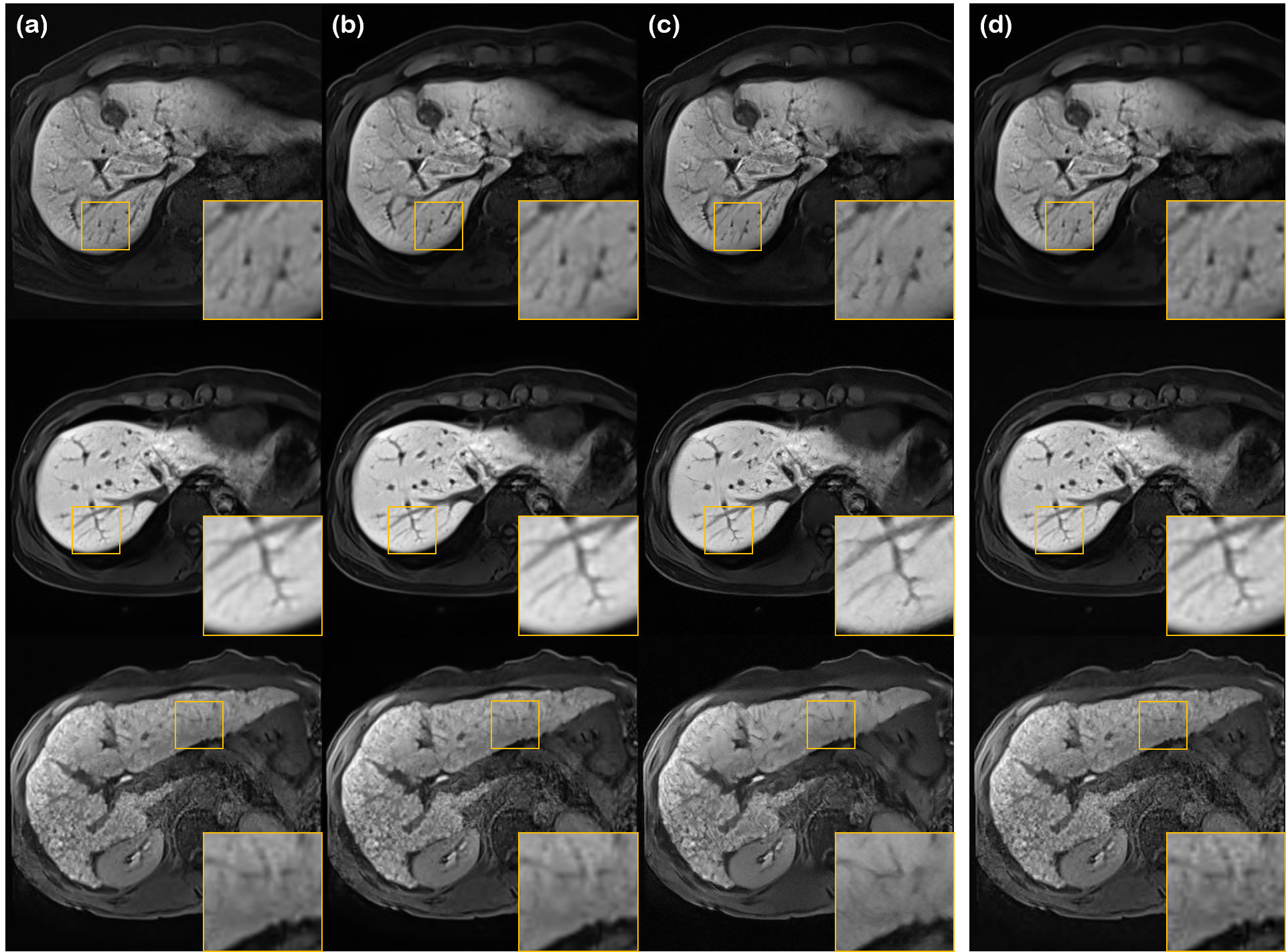}
    \caption{Denoising results using various methods on different liver conditions. (a) N2N~\cite{lehtinen2018noise2noise}, (b) N2Score~\cite{kim2021noise2score}, (c) proposed method, and (d) input noisy image. Yellow boxes show results that are magnified. First row: early liver cirrhosis with hepatocellular carcinoma (HCC), second row: normal liver, third row: advanced liver cirrhosis.}
	\label{fig:results_main}
\end{figure*}

\begin{figure*}[!hbt]
    \centering\includegraphics[width=18cm]{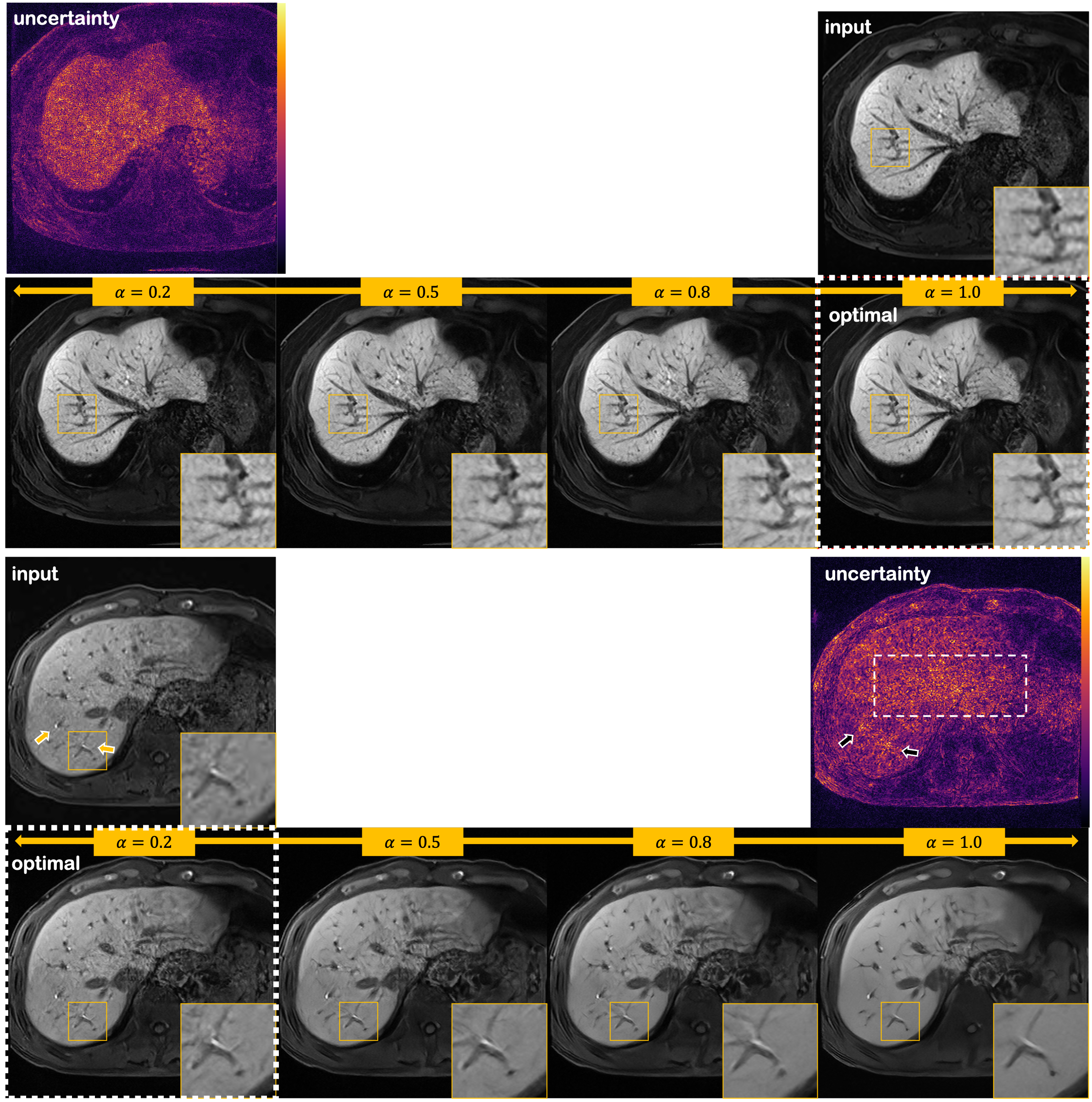}
    \caption{{Illustration of flexibility and uncertainty estimating property of the proposed method for the two patient data. 
    In the first row of each data, input noisy image and uncertainty map (standard deviation of the posterior mean) are shown.
    The second row illustrates denoised results with varying values of $\alpha$. The arrows indicate regions of high uncertainty, and the white box with dotted line points out region with concentrated noise. White boxes with the dotted line indicate the values of $\alpha$ used to compute the posterior mean.}}
	\label{fig:uncertainty}
\end{figure*}

\section{Results}
\label{sec:results}

\subsection{In vivo study}
\label{sec:result_real}

Denoising results are summarized in Fig.~\ref{fig:results_main}, with the quantitative metrics in TABLE~\ref{tab:snr_cnr}. In Fig.~\ref{fig:results_main}, we see that our method is the only method to effectively remove the noise while sharpening the image, {regardless of the given condition of the liver}. 
This is also consistent with the metrics given in TABLE~\ref{tab:snr_cnr}, where our method is the highest scoring method both in terms of SNR and CNR. Notably, it can be seen from TABLE~\ref{tab:snr_cnr} that while all methods have higher SNR than the input such that it is statistically relevant, R2D2+ is the only method which does not compromise CNR. 
This can be deciphered in two aspects - 1) R2D2+ is the only method that explicitly enhances the resolution with the same network that was used to denoise the image. 2) Posterior sampling does not produce blurry output, as opposed to MMSE estimates.

\begin{table}[!bt]
    \centering
    \begin{tabular}{c|c|c}
         & SNR & CNR \\ \hline
        Input & 15.28 $\pm$ 5.47 & \underline{14.46}$^{\dagger}$ $\pm$ 4.95 \\
        BM3D~\cite{dabov2006image} & 17.00$^{*}$ $\pm$ 6.29 & 11.82 $\pm$ 8.16 \\
        N2N~\cite{lehtinen2018noise2noise} & \underline{20.88}$^{*}$ $\pm$ 10.06 & 14.45$^{\dagger}$ $\pm$ 0.98 \\
        N2Score~\cite{kim2021noise2score} & 20.45$^{*}$ $\pm$ 8.08 & 14.27$^{\dagger}$ $\pm$ 9.56 \\
        Nei2Nei~\cite{huang2021neighbor2neighbor} & 16.69$^{*}$ $\pm$ 5.66 & 11.20 $\pm$ 7.37 \\
        R2D2+ & \textbf{21.28}$^{*}$ $\pm$ 8.35 & \textbf{16.08}$^{\dagger}$ $\pm$ 10.00 \\ \hline
    \end{tabular}
    \caption{Quantitative evaluation of SNR and CNR with different methods. Numbers in boldface indicate the best performing method, and numbers that are underlined indicate the second best. For further details, see TABLE~\ref{tab:snr_stats} and TABLE~\ref{tab:cnr_stats}.
    $^{*}: p < 0.0001$: pairwise comparison against Input.
    $^\dagger: p < 0.0001$: pairwise comparison against BM3D.
    Bonferroni correction applied~\cite{rupert2012simultaneous}.}
    \label{tab:snr_cnr}
\end{table}

From TABLE~\ref{tab:snr_cnr}, we do observe that self-supervised learning based methods (i.e., N2N, N2Score) have much better performance than BM3D. Nevertheless, the metrics lack far behind R2D2+, especially in terms of CNR. This can also be visually inspected in Fig.~\ref{fig:results_main} (b-d). While we can see relatively clearly that denoising has taken place as opposed to the noisy input, the denoised output tend to be washed out, with unclear boundaries and vessel structures. We note that Nei2Nei performs even worse than BM3D, which we conjecture to be stemming from its poor generalization capability. For further experimental results with detailed information about the statistical analysis, together with more figures including comparisons with BM3D, see Fig.~\ref{fig:results}.

\subsection{Flexibility and uncertainty}
\label{sec:flexibility}

{As was discussed earlier, we can flexibly adjust the parameter $\alpha$ so that we can get just the right amount of denoising that we want. An illustration of such control is shown in Fig.~\ref{fig:uncertainty}, where we vary the value of $\alpha$ from $0.2$ to $1.0$.} 

{In the first sample, it is clearly seen that we are able to achieve better results with higher values of $\alpha$. With a lower factor, the noise is insufficiently removed. This is the case where noise is distributed in a holistic manner, and hence the noise distribution is closer to the Gaussian distribution (see the input image). Since our noise estimator~\cite{chen2015efficient} is able to estimate the noise variance with high accuracy in this case, we do not have to alter the estimated value and simply set $\alpha$ close to $1.0$.}

{On the other hand, in the second sample, we see that with a high value of $\alpha$, the texture and the important details of the liver are washed out, and hence it would be better to keep the denoising factor $\alpha$ at a low level. Notice that pulsation artifact~\cite{budrys2018artifacts} is apparent in the central region of the image, also clearly seen in the input image. In such cases, the periphery of the image is rather clean, whereas the central region of the image is contaminated with very high scale of noise. Hence, it can be seen that the distribution of noise is far from Gaussian, leading to an over-estimation of the noise variance. Therefore, lower values of $\alpha$ may be needed. In any case, being able to control the amount of denoising is a prominent feature of our proposed method, and would be of great importance in clinical settings.}

Another important feature of the proposed method is the ability to quantify uncertainty. Denoising is an inverse problem, and hence there can be many feasible solutions. In order to embrace this fact, prior works have proposed several ways to estimate the uncertainty. For instance, \cite{laves2020uncertainty} leveraged MC dropout~\cite{gal2016dropout}. Other works tried to estimate the variance directly~\cite{kendall2017uncertainties}. In contrast, with the proposed method, we can simply sample multiple posterior samples, and directly compute the statistics from these samples.
{In order to show such property of R2D2+, we took 5 different posterior samples given the noisy image, and computed the mean and the standard deviation of the samples. The results are shown in the uncertainty maps of Fig.~\ref{fig:uncertainty}. The first sample does not depict any excessive variance in specific regions, and hence we can be fairly confident with the resulting denoised image. Contrarily, we do observe some regions of higher variance in the second sample (marked with arrows), which may indicate that a clinician should be conservative on decision making based on the denoised samples.}

\section{Discussion and related works}
\label{sec:discussion}

\subsection{Impact of low-frequency regularization}
\label{sec:lf_reg}

\begin{figure}[!h]
    \centering\includegraphics[width=8cm]{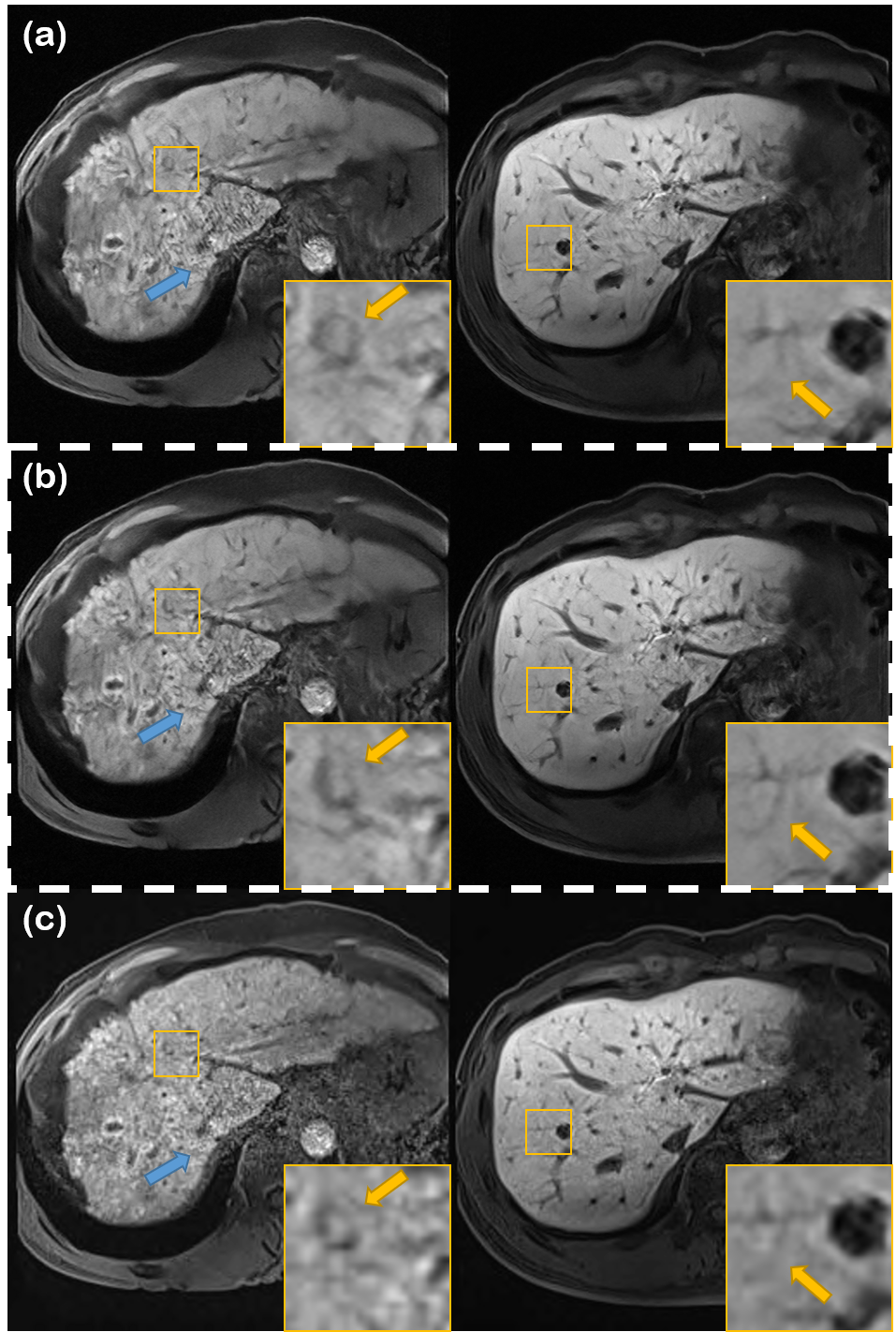}
    \caption{Ablation study on low-frequency regularization method. (a) Without regularization, (b) with regularization, and (c) noisy image. First column: The alteration of the tortuous vascular structure might be misinterpreted as a round hepatic focal lesion by radiologists in clinical practice. Blue arrows indicate the region where structured shaggy artifact is present in the reconstruction without regularization. Second column: clearer vessel structure can be seen in the yellow arrows. White dotted lines indicate the results by the proposed method.}
	\label{fig:lf_reg_ab}
\end{figure}

We study the effect of low-frequency regularization in Fig.~\ref{fig:lf_reg_ab}. In the first {column} (see yellow arrow), we see that the vessel structure is altered when the regularization is not performed. Contrarily, the structure is conserved with the regularization present. Furthermore, in the blue arrow of the first {column of}  Fig.~\ref{fig:lf_reg_ab}, we observe some shaggy-looking structured artifact when we do not perform any regularization. Again, this artifact is eliminated as we impose our regularization. Consistent with the observation that was made in the first {column}, we again see altered structure in the second {column}, when regularization is not imposed. Thus, we can safely conclude that the proposed regularization is able to lead our reconstructions toward a conservative outcome, thereby eliminating false positives/negatives.

\subsection{Ablation study on super-resolution}

\begin{figure}[!h]
    \centering\includegraphics[width=8cm]{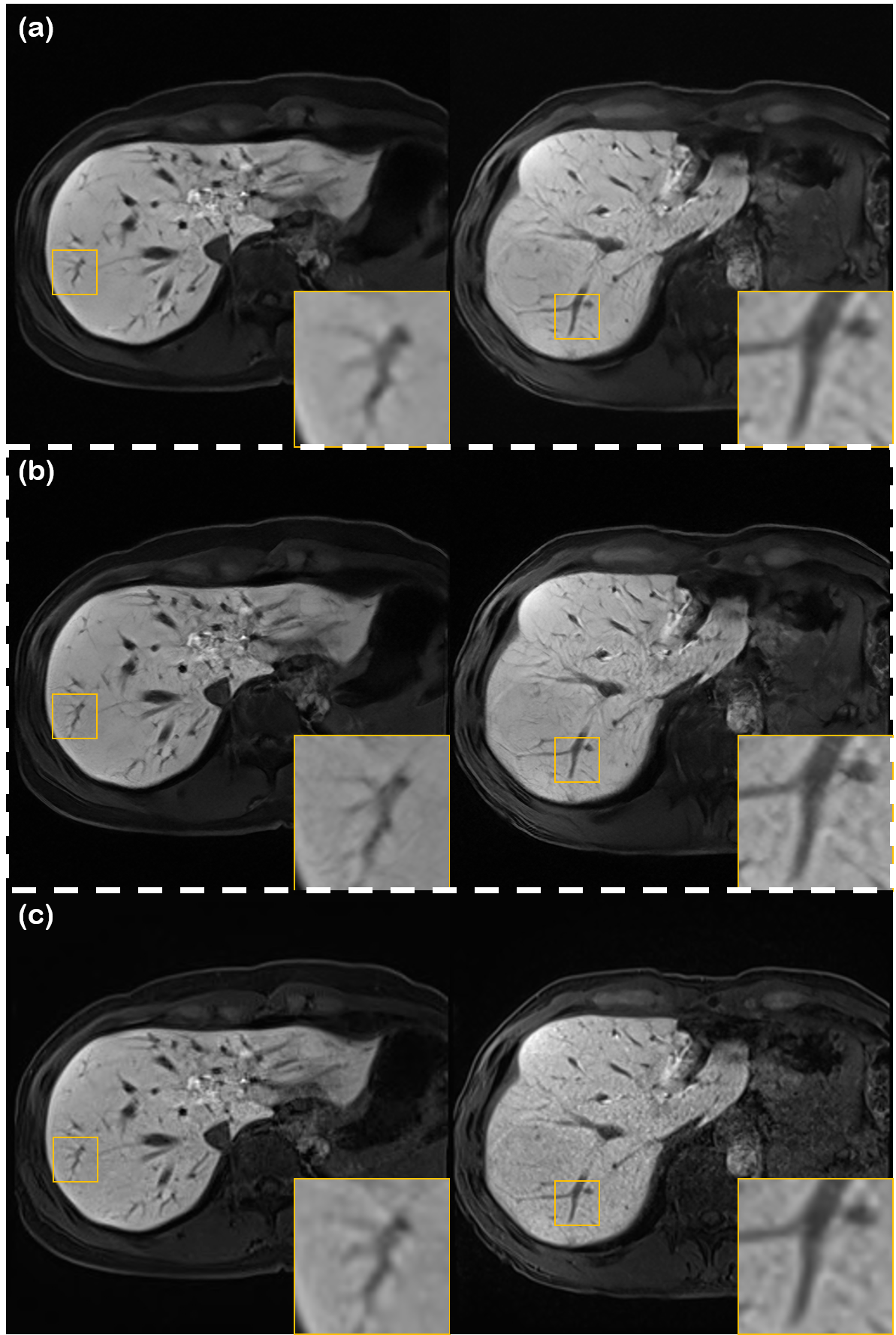}
    \caption{Ablation study on the post-hoc super-resolution method. (a) Without SR, (b) with SR, and (c) noisy image. White dotted lines indicate the results by the proposed method.}
	\label{fig:SR_ab}
\end{figure}

\begin{table}[!hbt]
    \centering
    \begin{tabular}{c|c|c}
         & SNR & CNR \\ \hline
        SR \cmark & 21.28 $\pm$ 8.35 & \textbf{16.08} $\pm$ 10.00 \\
        SR \xmark & \textbf{22.53} $\pm$ 9.06 & 11.24 $\pm$ 7.32 \\ \hline
    \end{tabular}
    \caption{Quantitative evaluation of SNR and CNR with/without resolution enhancement scheme.}
    \label{tab:SR_ab}
\end{table}

In Fig.~\ref{fig:SR_ab}, we compare the reconstruction results with and without using the post-hoc super-resolution method, with the corresponding metrics reported in TABLE~\ref{tab:SR_ab}. From the figure and the table, we see that the additional SR step greatly enhances the resolution, and the CNR, at the expense of a slight decrease in SNR. Hence, we can safely conclude that the proposed method has an overall positive effect on the reconstruction.

\subsection{Related works}
Denoising based on posterior sampling was discussed {in two of the recent workshop papers}~\cite{ohayon2021high, kawar2021stochastic}. Ohayon et al.~\cite{ohayon2021high} trains a conditional GAN (cGAN), which is trained in a retrospective supervised manner. At the inference stage, one can sample multiple different latent vectors $\zb$, so that one would be able to acquire multiple posterior samples given the same noisy image.
Kawar et al.~\cite{kawar2021stochastic} is the closest to our work, in that the authors propose to use a score-based generative model\footnote{Based on NCSNv2~\cite{song2020improved}}. However, there are several contributions that our paper makes over~\cite{kawar2021stochastic}. First, the prior work~\cite{kawar2021stochastic} uses un-regularized denoising steps, whereas ours use regularized steps based on the theory of stochastic contraction~\cite{chung2022come}. We have already shown in Section~\ref{sec:lf_reg} the superiority in preserving the content. Second, our method is fully unsupervised in that we use heavily out-of-distribution open source data to train our network, and validate our method on real-world {\em in vivo} data. In opposition,~\cite{kawar2021stochastic} only tests their denoising method on well-controlled in-distribution datasets. Finally, we utilize a more advanced form of score matching and sampler, which was verified to have a large gap against the previous work~\cite{song2020score}.
 
\section{Conclusion}
\label{sec:conclusion}

In this work, we proposed a denoising method for MRI, based on reverse diffusion and non-expansive mapping regularization. Along with it, we propose a method to super-resolve the denoised image with the {\em same} neural network. Our method achieves state-of-the-art both in terms of SNR and CNR, outperforming comparison methods by a large margin. With statistical analysis, we show that our method is the only method that can boost SNR such that it is statistically relevant, and at the same time boost the CNR. Furthermore, our method is able to quantify uncertainty in the solutions to the given inverse problem, while being flexible so that users can tweak how such noise should be eliminated, according to their needs. We believe that our research opens up an exciting new direction of denoising for medical imaging. 

\section*{Acknowledgment}
This work was supported by Institute of Information \& communications Technology Planning \& Evaluation (IITP) grant funded by the Korea government(MSIT) (No.2019-0-00075, Artificial Intelligence Graduate School Program(KAIST)), and by National Research Foundation(NRF) of Korea grant NRF-2020R1A2B5B03001980.

\section*{Appendix}

\section*{Statistical Analysis}

Detailed results from performing RM-ANOVA to SNR/CNR values can be seen in TABLE~\ref{tab:snr_stats}, and TABLE~\ref{tab:cnr_stats}, respectively.

\section*{Additional experimental results}

\begin{figure*}[t]
    \centering\includegraphics[width=17cm]{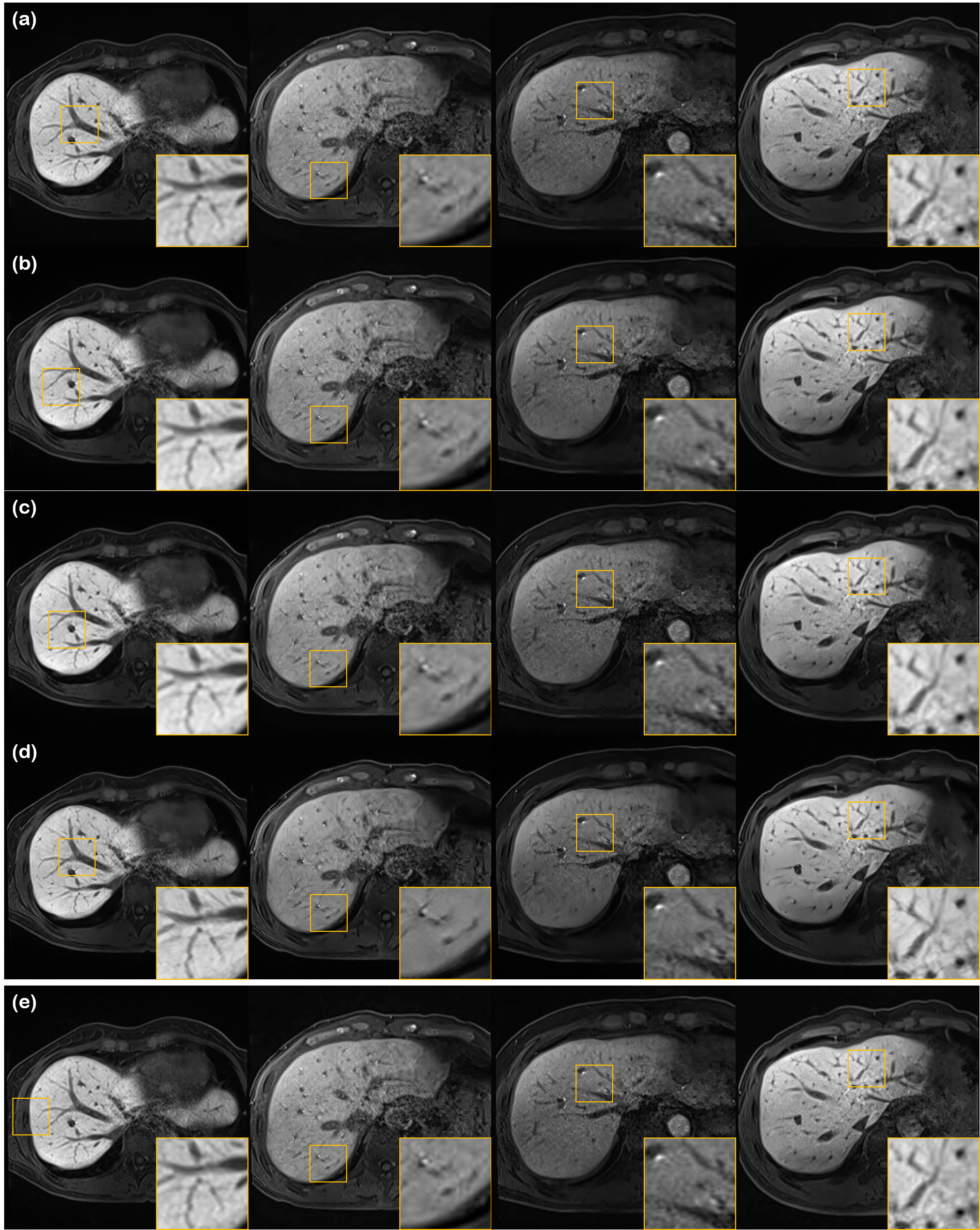}
    \caption{{Denoising results using various methods. (a) BM3D~\cite{dabov2006image}, (b) N2N~\cite{lehtinen2018noise2noise}, (c) N2Score~\cite{kim2021noise2score}, (d) proposed method, and (e) Input noisy image. Yellow boxes show results that are magnified.}}
	\label{fig:results}
\end{figure*}

In Fig.~\ref{fig:results}, we show additional experimental results, comparing extensively with block-matching 3D filtering (BM3D)~\cite{dabov2006image}, Noise2Noise (N2N)~\cite{lehtinen2018noise2noise}, and Noise2Score (N2Score)~\cite{kim2021noise2score}. It is clear from Fig.~\ref{fig:results} (d) that the proposed method clearly outperforms all methods, by reconstructing sharp vessel structures, while removing excessive noise around such structure. All other comparison methods fall heavily behind the proposed method, especially in terms of preserving high-frequency detail.

\ifCLASSOPTIONcaptionsoff
  \newpage
\fi

\bibliographystyle{IEEEtran}
\bibliography{reference}

\clearpage
\newpage

\begin{table}[t]
    \centering
    \begin{tabular}{c|c|c|c|c}
        Method 1 & Method 2 & Mean diff. & Std. Error & $p^{*}$ \\
        \hline
        \multirow{5}{*}{Input} & BM3D & -1.72 & 0.108 & \textbf{$<$0.0001} \\
        & Nei2Nei & -1.41 & 0.086 & \textbf{$<$0.0001} \\
        & N2N & -5.60 & 0.30 & \textbf{$<$0.0001} \\
        & N2Score & -5.17 & 0.25 & \textbf{$<$0.0001} \\
        & proposed & -5.99 & 0.32 & \textbf{$<$0.0001} \\ \hline
        \multirow{5}{*}{BM3D} & input & 1.72 & 0.11 & \textbf{$<$0.0001} \\
        & Nei2Nei & 0.31 & 0.085 & 0.0055 \\
        & N2N & -3.87 & 0.24 & \textbf{$<$0.0001} \\
        & N2Score & -3.45 & 0.19 & \textbf{$<$0.0001} \\
        & proposed & -4.27 & 0.31 & \textbf{$<$0.0001} \\ \hline
        \multirow{5}{*}{Nei2Nei} & input & 1.41 & 0.086 & \textbf{$<$0.0001} \\
        & BM3D & -0.31 & 0.085 & 0.0055 \\
        & N2N & -4.18 & 0.27 & \textbf{$<$0.0001} \\
        & N2Score & -3.76 & 0.23 & \textbf{$<$0.0001} \\
        & proposed & -4.58 & 0.32 & \textbf{$<$0.0001} \\ \hline
        \multirow{5}{*}{N2N} & input & 5.60 & 0.30 & \textbf{$<$0.0001} \\
        & BM3D & 3.87 & 0.24 & \textbf{$<$0.0001} \\
        & Nei2Nei & 4.18 & 0.27 & \textbf{$<$0.0001} \\
        & N2Score & 0.42 & 0.11 & 0.0040 \\
        & proposed & -0.40 & 0.40 & 1.000 \\ \hline
        \multirow{5}{*}{N2Score} & input & 5.17 & 0.25 & \textbf{$<$0.0001} \\
        & BM3D & 3.45 & 0.19 & \textbf{$<$0.0001} \\
        & Nei2Nei & 3.76 & 0.22 & \textbf{$<$0.0001} \\
        & N2N & -0.42 & 0.11 & 0.0040 \\
        & proposed & -0.82 & 0.38 & 0.51 \\ \hline
        \multirow{5}{*}{proposed} & input & 5.99 & 0.32 & \textbf{$<$0.0001} \\
        & BM3D & 4.27 & 0.31 & \textbf{$<$0.0001} \\
        & Nei2Nei & 4.58 & 0.32 & \textbf{$<$0.0001} \\
        & N2N & 0.40 & 0.40 & 1.000 \\
        & N2Score & 0.82 & 0.38 & 0.5097 \\ \hline
    \end{tabular}
    \caption{Details of pairwise comparison results in RM ANOVA statistics of SNR.
    $^{*}$: Bonferroni corrected.}
    \label{tab:snr_stats}
\end{table}

\begin{table}[t]
    \centering
    \begin{tabular}{c|c|c|c|c}
        Method 1 & Method 2 & Mean diff. & Std. Error & $p^{*}$ \\
        \hline
        \multirow{5}{*}{Input} & BM3D & 2.62 & 1.04 & 0.223 \\
        & Nei2Nei & 3.24 & 0.91 & 0.021 \\
        & N2N & -0.016 & 1.27 & 1.000 \\
        & N2Score & 0.175 & 1.22 & 1.000 \\
        & proposed & -1.62 & 1.30 & 1.000 \\ \hline
        \multirow{5}{*}{BM3D} & input & -2.62 & 1.04 & 0.223 \\
        & Nei2Nei & 0.62 & 0.15 & 0.0022 \\
        & N2N & -2.64 & 0.34 & \textbf{$<$0.0001} \\
        & N2Score & -2.45 & 0.26 & \textbf{$<$0.0001} \\
        & proposed & -4.25 & 0.63 & \textbf{$<$0.0001} \\ \hline
        \multirow{5}{*}{Nei2Nei} & input & -3.24 & 0.96 & 0.0211 \\
        & BM3D & -0.62 & 0.15 & 0.0022 \\
        & N2N & -3.26 & 0.44 & \textbf{$<$0.0001} \\
        & N2Score & -3.07 & 0.36 & \textbf{$<$0.0001} \\
        & proposed & -4.87 & 0.65 & \textbf{$<$0.0001} \\ \hline
        \multirow{5}{*}{N2N} & input & 0.016 & 1.24 & 1.000 \\
        & BM3D & 2.64 & 0.34 & \textbf{$<$0.0001} \\
        & Nei2Nei & 3.26 & 0.43 & 1.000 \\
        & N2Score & 0.19 & 0.19 & 0.493 \\
        & proposed & -1.61 & 0.73 & 1.000 \\ \hline
        \multirow{5}{*}{N2Score} & input & -0.17 & 1.21 & 1.000 \\
        & BM3D & 2.45 & 0.26 & \textbf{$<$0.0001} \\
        & Nei2Nei & 3.07 & 0.36 & \textbf{$<$0.0001} \\
        & N2N & -0.19 & 0.19 & 1.000 \\
        & proposed & -1.80 & 0.67 & 0.149 \\ \hline
        \multirow{5}{*}{proposed} & input & 1.62 & 1.30 & 1.000 \\
        & BM3D & 4.25 & 0.63 & \textbf{$<$0.0001} \\
        & Nei2Nei & 4.87 & 0.65 & \textbf{$<$0.0001} \\
        & N2N & 1.61 & 0.73 & 0.493 \\
        & N2Score & 1.80 & 0.67 & 0.149 \\ \hline
    \end{tabular}
    \caption{Details of pairwise comparison results in RM ANOVA statistics of CNR.
    $^{*}$: Bonferroni corrected.}
    \label{tab:cnr_stats}
\end{table}

\end{document}